**Title:** Bias in the estimation of cumulative viremia in cohort studies of HIV-infected individuals

**Authors:** Maia Lesosky[a], Tracy Glass[a], Brian Rambau[a], Nei-Yuan Hsiao[b], Elaine J. Abrams[c], Landon Myer[a]

**Affiliations:**

[a] Division of Epidemiology & Biostatistics, School of Public Health & Family Medicine, University of Cape Town, Observatory 7925, South Africa

[b] Division of Medical Virology, Department of Pathology, University of Cape Town and National Health Laboratory Services, Observatory 7925, South Africa

[c] ICAP at Columbia University, Mailman School of Public Health, Columbia University and College of Physicians and Surgeons, Columbia University, Columbia University, New York, New York, United States of America

Email address:

TG: tracy.glass@uct.ac.za

BR: brian.rambau@uct.ac.za

N-YH: marvin.hsiao@uct.ac.za

EJA: eja1@columbia.edu

LM: landon.myer@uct.ac.za

*Corresponding author: Maia Lesosky, lesosky@gmail.com/maia.lesosky@uct.ac.za; +27 21 650 4532, School of Public Health & Family Medicine, Health Sciences Campus, University of Cape Town, Private Bag X3, Rondebosch 7701, South Africa

**Abstract**

**Purpose:** The use of cumulative measures of exposure to raised HIV viral load (viremia copy-years) is an increasingly common in HIV prevention and treatment epidemiology due to the high biological plausibility. We sought to estimate the magnitude and direction of bias in a cumulative measure of viremia caused by different frequency of sampling and duration of follow-up.

**Methods:** We simulated longitudinal viral load measures and reanalysed cohort study datasets with longitudinal viral load measurements under different sampling strategies to estimate cumulative viremia.

**Results:** In both simulated and observed data, estimates of cumulative viremia by the trapezoidal rule show systematic upward bias when there are fewer sampling time points and/or increased duration between sampling time points, compared to estimation of full time series. Absolute values of cumulative viremia vary appreciably by the patterns of viral load over time, even after adjustment for total duration of follow up.

**Conclusions:** Sampling bias due to differential frequency of sampling appears extensive and of meaningful magnitude in measures of cumulative viremia. Cumulative measures of viremia should be used only in studies with sufficient frequency of viral load measures and always as relative measures.

# 1. Introduction

Cumulative measures of exposure are widely used in different areas of epidemiologic research in the estimation of the effects associated with exposures over time [1]. In infectious diseases epidemiology, cumulative viral load (cVL), or viremia copy-years, has been proposed as an estimate of the cumulative exposure to raised HIV-RNA viral load (VL) in individuals infected with human immunodeficiency virus (HIV). This cumulative exposure measure has high biological plausibility: raised VL is the primary marker of both disease control and infection transmission risk [2]. In individuals receiving antiretroviral therapy (ART), raised VL is associated with treatment failure and poor long-term treatment outcomes [3]. Cumulative exposure to raised VL has been hypothesised as a potential risk factor, or predictor of poor outcomes [4-6].

cVL appears to have been introduced independently by Zoufaly et al. [7] and Cole et al. [8] in association with outcomes related to the development of acquired immune deficiency syndrome (AIDS). Subsequently, variants of cVL appeared in use across a variety of studies, where the measure was often intended as a potential risk factor to predict outcomes such as all-cause mortality [4-6,9,10]. The empirical evidence for cVL as a useful prognostic factor for long term outcomes is mixed. While there is a biological argument to be made that the cumulative exposure to raised VL [11, 12] could be predictive of health outcomes, the empirical measurement of cumulative exposure by cVL has failed to demonstrate consistent associations. For example, Cates et al [13] conclude that cVL burden prior to pregnancy is not informative in a study of risk of pregnancy loss among HIV infected women, but current VL appears highly predictive. In a follow up efficacy study comparing the antiretroviral drugs efavirenz and boosted lopinavir, cautious optimism was expressed towards the use of cVL as a prognostic factor [14]. Kukoyi et al [15] found that a log10 copies/mL/year measure to evaluate morbidity outcomes in a paediatric cohort was predictive, but the VL measure is included in analyses as a discrete variable, with just three levels. Kowalkowski et al [16] also found associations in sub-analyses between a measure of cVL and non-AIDS defining malignancies, utilising a complex series of data imputation on 7-day follow up windows. Taken together, it remains unclear if cVL has greater utility then current or most recent VL.

Common concerns related to cumulative measures of exposure in epidemiology include time-based dependence of exposure and the method of measurement (direct or indirect) [7]. cVL may be defined as the area under the plasma VL curve and has been calculated using copies/mL or $\log_{10}$ copies/mL. However, the phrase "under the curve" is not precise, as VL is not continuously measured over time, but sampled at discrete time points, and therefore the true individual trajectory is unknown and the area must be approximated using integral approximation methods. Some of the mixed empirical results in estimating the association of cVL to health outcomes may be due to the inconsistencies in estimation. The majority of analyses estimate cumulative viremia or viral copy-years using the trapezoidal rule, which sums the areas between two successive points joined by linear interpolation [17]. While it has been established that this method is scale dependent [18], no research has examined the impact of sampling frequency bias. Here we estimate the bias due to changes in sampling frequency when estimating cVL by the trapezoidal rule. We utilise two empirical and one simulated data set, each with repeated VL measures, to estimate the magnitude and direction of the bias in the estimation of cVL and make recommendations for improved usage.

## 2. Methods

*2.1 Data sources*

We used three different data sources to examine how sampling time points effect cVL estimation: an intensively monitored cohort of women initiating ART [11], a simulated data set of longitudinal HIV VL trajectories developed for a monitoring study [19], and a random sample of routine VL data collected by the South African National Health Laboratory Services (NHLS) [20]. The two empirical datasets were selected to reflect different types of data collection, either under intensive health research and follow-up or under routine care and with different expected longitudinal VL trajectories drawing from different populations. The inclusion of simulated data provides an opportunity to evaluate estimation of cVL at different sampling frequencies against a known 'truth'.

The cohort study has VL measures of 518 women initiating ART during pregnancy and followed up to approximately 18 months postpartum. This is a frequently measured cohort with a median of 7 VL measures during the follow up, which is much more frequent than most routine clinical settings. The routine monitoring sample includes n = 29,519 individuals, who were not virally suppressed for the entire duration of follow-up and had a minimum of 5 VL measures from the NHLS data set. The sample is extracted from routine laboratory data comprised of individuals in ART programmes in the Western Cape, South Africa. This routinely collected monitoring data has either 6 or 12 months between VL measures, on average, following South African treatment guidelines [21]. The simulated dataset is contributed from a model of longitudinal VL trajectories for 80 weeks on a weekly time step [20]. The data is simulated to mimic trajectories and conditions of patients initiating ART during and contains a simulated sample of 10,000 individuals.

This study is an analysis of secondary data which was originally collected by studies carried out under ethics approval from the University of Cape Town Faculty of Health Sciences Human Research Ethics Committee (HREC: 865/2016; 329/2014; 451/2012).

*2.2 Estimation of cVL*

$Log_{10}$ VL measures are used throughout in the calculation of cVL. cVL is calculated for each time series using the trapezoidal rule which uses a linear approximation between successive points (Equation 1 and Figure 1). VL measures below 50 copies per mL are set to zero to avoid an increase in cVL while below the limit of detection. In addition to estimating cVL over the given time frame (between first and last VL observations, regardless of total duration of observation) we calculate a standardised cVL that adjusts for individual follow-up time (Equation 2). Median (interquartile) range of the individual estimates of cVL are reported.

$$cVL = \frac{t_n - t_0}{n} \left[ \frac{VL(t_0) + VL(t_n)}{2} \sum_{k=1}^{n-1} VL(t_0 + k \frac{t_n - t_0}{n}) \right] \quad (1)$$

$$cVL_{FU} = cVL(J) / t_n \quad (2)$$

*2.3 Subsampling*

We calculate cVL based on subsets of the available data and compare these estimates to the $cVL_{REF}$, which uses every data point in the full time series. We use two different subsampling strategies: a) random subsample to a set number of VL measures (n=2 - 5), ensuring that the first and last VL measure per individual are always retained to ensure consistency with total follow up time and b) sampling based on average interval between VL measures, retaining the first and last VL measure per individual. Median and interquartile ranges for estimated cVL are provided, as well as the percent change compared to $cVL_{REF}$. This estimate reflects both the viremia and duration of follow up of individuals in the respective data sets. A follow-up standardised measure is also calculated and reported in order to reduce the dependence on the individual's total duration of follow-up. This measure is analogous to the viral copy-year. The $cVL_{FU}$ is calculated using all available VL measures, and then individually standardised for total follow-up time. Cohort viral load trajectories were visualised using generalised additive model smoothing with a natural b-spline.

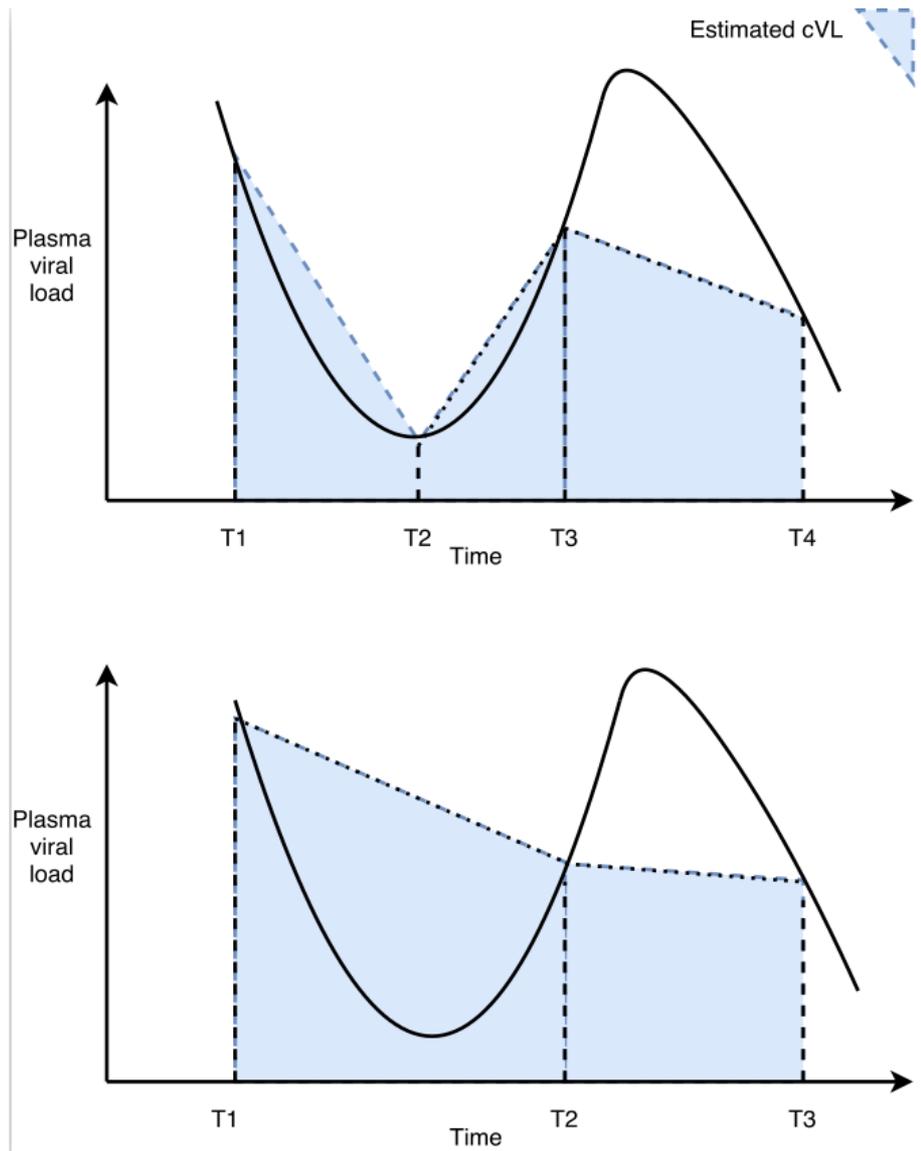

**Figure 1:** Schematic of plasma VL versus time for one profile under two different monitoring schemes. Profile A (top) has been measured at 4 time points and Profile B (bottom) at only three time points. In each profile the resulting estimated cVL is represented by the shaded region.

## 3 Results

Characteristics of the three data sets can be found in Table 1. The data sources vary over indicators including the median number of VL measures available per individual, the median time between VL measures and the total duration of follow up. The median duration between VL measures varies from a low of one week (simulated data) to a high of five months (routine laboratory data). The median VL measures available per person is 7 (IQR: 5-9), 7 (IQR: 6-9) and 21 (IQR: 15-26) in the cohort, routine monitoring and in the simulated data sets respectively.

The duration of follow up is the longest in the routine data, at a median of 3.8 years (IQR: 2.7-5.2 years) and shortest in the simulated VL data at 0.4 years (IQR: 0.3, 0.5 years).

**Table 1:** Summary characteristics of data sets used for estimation of cVL

|  | **Simulated data** n =10,000 | **Cohort** n = 518 | **Routine laboratory** n = 29,519 |
|---|---|---|---|
| VL measures (N) | 601,070 | 3,519 | 253,503 |
| Proportion VL < 50 | 0.17 | 0.66 | 0.39 |
| Proportion VL 50-1000 | 0.67 | 0.16 | 0.188 |
| Proportion VL >1000 | 0.16 | 0.19 | 0.43 |
| VL measures available per individual median (Q1, Q3) | 61 (55, 65) | 7 (5, 9) | 7 (6, 9) |
| Follow-up duration per individual (years) median (Q1, Q3) | 1.2 (1.0, 1.2) | 1.1 (0.6, 1.4) | 3.8 (2.7, 5.2) |
| Follow-up duration per individual (weeks) median (Q1, Q3) | 60.0 (54.0, 64.0) | 59.6 (33.6, 71.6) | 199.7 (142.4, 271.7) |
| Median time between observations per individual median (Q1, Q3) (weeks) | 1.0 (1.0,1.0) | 7.1 (4.9, 13.0) | 19.8 (4.0, 33.7) |
| Total person-years | 11,367 | 548 | 117,279 |

The cVL$_{REF}$ is calculated using all available VL measures in each data set and represents the most accurate possible measure of cVL (Table 2). The median cVL measure is 0.49 log$_{10}$ copies/mL years (IQR: 0.20-0.97 log$_{10}$ copies/mL years) using the cohort data, 7.39 log$_{10}$ copies/mL years (IQR: 4.55-11.41 log$_{10}$ copies/mL years) in the routine monitoring laboratory data, and 2.08 log$_{10}$ copies/mL years (IQR: 1.84-2.72 log$_{10}$ copies/mL years) in the simulated data. The median standardised cVL measure in the cohort is 0.50 log$_{10}$ copies/mL year (IQR: 0.23-1.34 log$_{10}$ copies/mL year), in the routine laboratory data 2.10 log$_{10}$ copies/mL year (IQR: 1.29-3.12 log$_{10}$ copies/mL year), and in the simulated data 2.09 log$_{10}$ copies/mL year (IQR: 1.57-2.24 log$_{10}$ copies/mL year).

**Table 2**: Estimated cVL using complete time series per individual (bold) and using different random sub-samples of observations. Estimates given as median (IQR) of individual cVL calculations and as relative percent change between median cVL levels. Values given as median (Q1, Q3) unless specified.

|  | **Simulated data** | **Cohort** | **Routine laboratory** |
|---|---|---|---|
| **Reference cVL ($cVL_{REF}$)** | **2.08 (1.84, 2.72)** | **0.49 (0.20, 0.97)** | **7.39 (4.55, 11.41)** |
| $cVL_2$ subsample: N = 2 VL/person (first and last observations) | 2.99 (2.44, 3.71) | 2.17 (0.93, 3.18) | 7.52 (3.98, 11.82) |
| Percent change from $cVL_{REF}$ | +44% | +344% | +2% |
| $cVL_3$ subsample: N = 3 VL/person | 2.62 (2.08, 3.24) | 1.05 (0.53, 1.93) | 7.69 (4.35, 11.94) |
| Percent change from $cVL_{REF}$ | +26% | +115% | +4% |
| $cVL_4$ subsample: N = 4 VL/person | 2.48 (1.98, 3.06) | 0.79 (0.42, 1.53) | 7.58 (4.43, 11.90) |
| Percent change from $cVL_{REF}$ | +19% | +62% | +3% |
| $cVL_5$ subsample: N = 5 VL/person | 2.41 (1.94, 2.95) | 0.73 (0.37, 1.42) | 7.52 (4.46, 11.82) |
| Percent change from $cVL_{REF}$ | +16% | +50% | +2% |
| **Standardised reference cVL ($cVL_{FUD} = cVL_{REF}$ / FUD_yrs)** | **2.09 (1.57, 2.24)** | **0.50 (0.23, 1.34)** | **2.1 (1.29, 3.12)** |
| $cVL_2$ subsample: N = 2 VL/person (first and last observations) | 2.65 (2.19, 3.22) | 2 (1.54, 2.50) | 2.21 (1.27, 3.14) |
| Percent change from $cVL_{FUD}$ | +27% | +301% | +5% |
| $cVL_3$ subsample: N = 3 VL/person | 2.34 (1.86, 2.84) | 1.24 (0.59, 1.74) | 2.21 (1.35, 3.21) |
| Percent change from $cVL_{FUD}$ | +12% | +149% | +5% |
| $cVL_4$ subsample: N = 4 VL/person | 2.24 (1.78, 2.65) | 0.79 (0.37, 1.50) | 2.18 (1.33, 3.21) |
| Percent change from $cVL_{FUD}$ | +7% | +59% | +4% |
| $cVL_5$ subsample: N = 5 VL/person | 2.20 (1.72, 2.55) | 0.59 (0.30, 1.26) | 2.15 (1.32, 3.19) |
| Percent change from $cVL_{FUD}$ | +5% | +18% | +2% |

Meaningfully large differences in estimated cVL are found between $cVL_{REF}$ compared to random subsamples with reduced numbers of observations (Table 2). There is a significant and meaningful bias towards overestimation, ranging from +50% to +344% change compared to cVL reference in the cohort, and a +16% to +44% change from reference in the simulated data, both of which reflect short duration (<2 years) follow-up. The routine monitoring data demonstrates a different pattern of bias of small but consistent overestimation of between +3% to +4% from reference cVL. This consistency is partially reflective of the high proportion (39%) of observations that are <50 copies/mL and the overestimation increases when the proportion of VL measures <50 copies/mL decreases. Use of the follow-up standardised reference measure reduces the extent of the bias in the cohort and simulation data, where it ranges from +18% to +301% (cohort) and +5% to +27% (simulation). In the routine laboratory data the change from reference remained consistently positive and ranged from +2% to +5%. In general, the different subsamples give estimates further away from the reference measure as the number of subsamples decreases (Table 2).

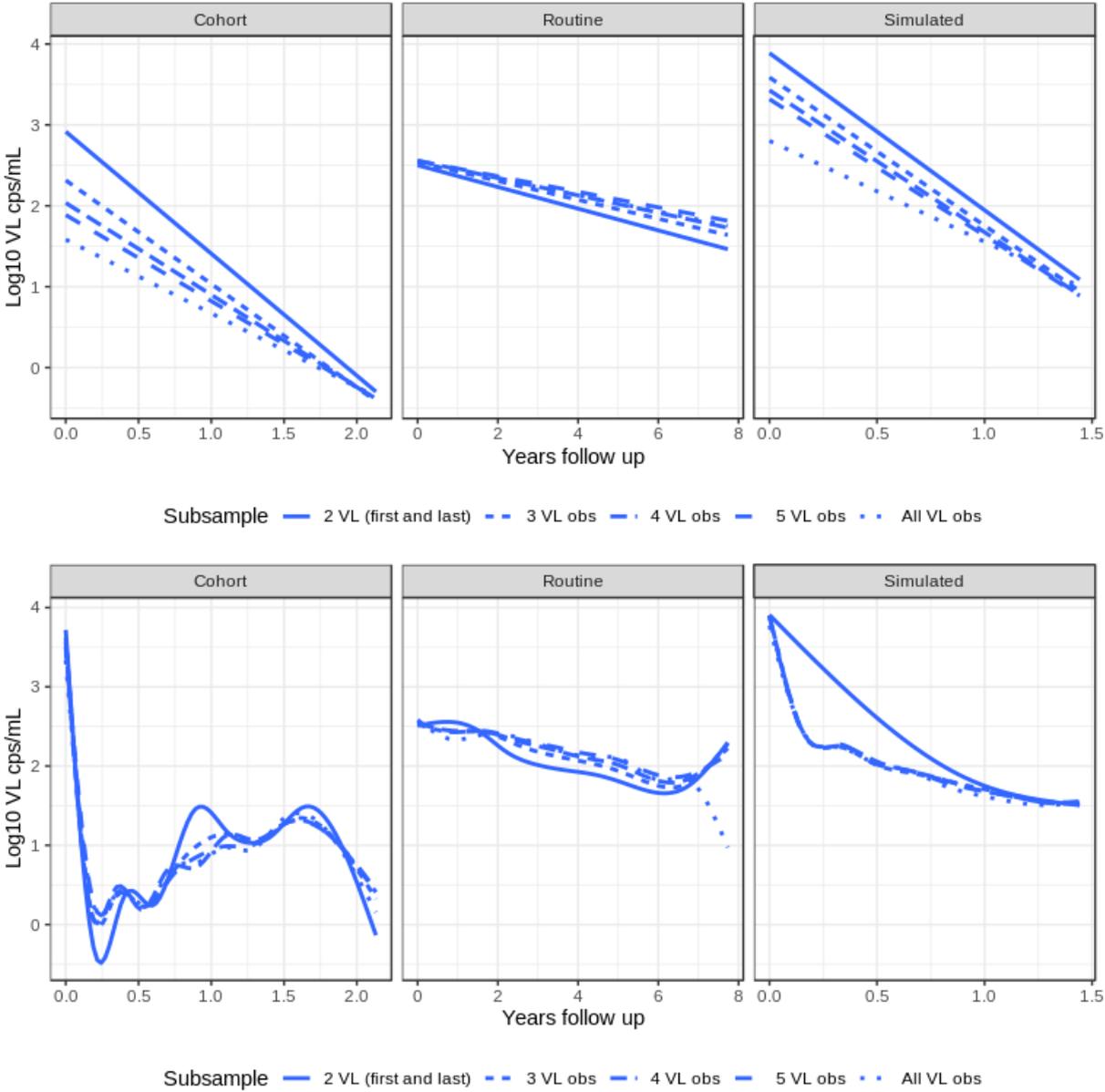

**Figure 2:** Smoothed sample trajectories for each data sets under sub-sampling. Top panel is linear fit and bottom panel fit by generalised additive models (GAMs) using all VL measures available in each of the available (subsampled) data sets.

Figure 2 shows smoothed trajectories for subsamples, under linear and generalised additive model smoothing, to aid in understanding the mean data trends under each of the subsampling strategies. Specifically we see that both the linear and spline fits show the 2 VL subsample superior to the fits of all other samples, including the complete data case in the cohort and simulated data, but this is reversed in the routine laboratory data.

When evaluating the subsamples where sampling was based on increasing the mean duration between VL measures, there was an increasing positive bias in the estimation of cVL, from +7% to +20% as the duration between samples increases from 4 weeks to 24 weeks. The empirical cohort data demonstrates the same trend, but with much larger bias, ranging from +36% at 4 weeks minimum duration to +335% at 24 weeks minimum duration between samples. It is important to note that the sample size reduces as the duration between samples increases in the cohort data. Estimation of the follow-up time standardised version of cVL for the same subsets indicates a similar tendency, but with a slightly smaller magnitude of bias compared to the non-standardised estimates, ranging to only +9% in the simulated data to +299% in the cohort data (Table 3).

**Table 3**: Estimated cVL using complete time series per individual (bold) and using different frequencies of observations for sub-samples. Estimates given as median (IQR) of individual cVL calculations and as relative percent change between median cVL levels. Values given as median (Q1, Q3) unless specified.

|  | Simulated data | | Cohort data | |
|---|---|---|---|---|
|  | **cVL$_{REF}$** | **cVL$_{FUD}$** | **cVL$_{REF}$** | **cVL$_{FUD}$** |
| **Reference cVL** (all samples) | **2.08 (1.84, 2.72)** | **2.09 (1.57, 2.24)** | **0.49 (0.20, 0.97)** | **0.50 (0.23, 1.34)** |
| 4 wks between samples | 2.25 (1.8, 2.74) | 2.07 (1.64, 2.31) | 0.67 (0.37, 1.12) | 0.73 (0.36, 1.52) |
| Percent change from cVL$_{REF}$ | +8% | -1% | +36% | +46% |
| 8 wks between samples | 2.23 (1.83, 2.75) | 2.07 (1.61, 2.34) | 0.79 (0.48, 1.55) | 1.07 (0.47, 1.83) |
| Percent change from cVL$_{REF}$ | +7% | +0% | +62% | +115% |
| 12 wks between samples | 2.31 (1.91, 2.82) | 2.14 (1.65, 2.4) | 0.98 (0.58, 1.93) | 1.26 (0.63, 1.95) |
| Percent change from cVL$_{REF}$ | +11% | +2% | +100% | +154% |
| 24 wks between samples | 2.5 (1.99, 3.02) | 2.27 (1.75, 2.61) | 2.12 (0.92, 3.09) | 1.99 (1.53, 2.48) |
| Percent change from cVL$_{REF}$ | +20% | +9% | +335% | +299% |

## 4 Discussion

Estimating the true cumulative exposure by approximating the area under a curve using repeat measures of a biomarker such as VL on time scales that are routine in clinical care (weeks to months) is subject to a number of sources of bias. As the "true" shape of the trajectory between measurements is unknown, approximation of cVL has typically been by a crude method of geometric approximation known as the trapezoidal rule. In most empirical work with repeated measures of VL, the sparsity of VL measures over time ensures that accurate approximation of the true trajectory is unlikely. Even within a single study, unless VL measures occur with the same timing and frequency for every individual, the measure of cVL may be more reflective of sampling frequency bias than any underlying biological mechanism: cVL for individuals with fewer, or more widely spaced measures will be inaccurate and biased upwards compared to individuals with more, or more frequent, VL measures.

Estimation of cVL via the trapezoidal rule has the potential for significant bias and inaccuracy depending on the frequency of VL measures and underlying stochastic process. In populations with moderate to high rates of partial or non-adherence, long periods between observations may result in inaccuracy with respect to estimation of the time an individual spends with elevated VL. Similarly, different frequencies of observation have a clear impact on estimated cumulative viremia and tend to, if anything, overestimate the exposure. Cumulative VL may be a useful indicator for relative exposure *within* studies if individuals have been subject to similar frequency and timing of viral load measures. Comparison of cVL between data sets or studies are best undertaken with a great deal of caution, as variation in observation frequency or timing invalidates comparison of the relative measure. While relative measures of exposure, for example, pack-years in smoking exposure, are routinely compared across studies, it is necessary to do so with an assumption that the underlying exposure is not changing substantially over time, though this is rarely stated.

Some approaches to reducing the impact of sampling bias have been applied. Mugavero et al. [4] use inverse weights to adjust for sampling frequency, resulting in a positive association between cVL and mortality outcomes. However, this approach is only applicable in the case where a reasonable approach to estimation of the inverse weights is possible. Other integral estimation methods are available, for example, composite Simpson's rule [17] which uses a third order polynomial to interpolate between successive points, and is know to be more accurate than trapezoidal estimation [17], but requires evenly spaced time points which are uncommon in most cohort studies. Another approach may be to avoid attempts at cumulative viraemia exposures altogether in favour of simpler measures which may be more directly interpretable [22]. Finally, we note that the use of the term cumulative viremia or viral-copy-years is non-precise, as different researchers have utilised slightly different definitions, both in terms of using copies/mL vs $\log_{10}$ copies/mL, different handling of VL measures below the limit of detection and different methods of follow-up duration standardisation. The key point here is that without standardisation of the estimate, and without an explicit process for managing sampling frequency and other sources of obvious bias, the measures of cumulative exposure to raised HIV viral load cannot be presumed consistent and are unlikely to be comparable across studies.

In summary, cVL exposure estimates should be used with caution as they will tend to overestimate the true exposure to raised viremia, and should not be used as an absolute measure of comparison between studies or cohorts with different frequencies of VL monitoring. Utilisation of cVL as a relative measure within a well characterised and consistently observed cohort may provide useful relative estimates of differential exposure to raised viral load over time.

**Acknowledgements** The authors thank the MCH-ART study and South African National Health Laboratory Service project for contribution of anonymized data utilised in this analysis.

**Competing interests** The authors have no competing interests to declare.

**Funding** The work was supported by the Eunice Kennedy Shriver National Institute of Child Health & Human Development of the National Institutes of Health [R21HD093463]. The content is solely the responsibility of the authors and does not necessarily represent the official views of the National Institutes of Health and by the Medical Research Council of South Africa in terms of the National Health Scholars Programme from funds provided for this purpose by the National Department of Health/Public Health Enhancement Fund; South African Department of Science and Technology/National Research Foundation (DST-NRF), Centre of Excellence in Epidemiological Modelling and Analysis (SACEMA), Stellenbosch University, Stellenbosch, South Africa to TG and BR.

**Authorship** ML and LM conceived of the study idea. ML, LM, EA, N-YH contributed to the acquisition of the data. ML, TG developed the simulation model. ML, BR, TG carried out the analysis and interpretation of the results. ML, TG wrote the first draft. All authors reviewed and critically revised the draft and approved the final version to be submitted.

**References**

[1] White E, Armstrong BK, Saracci R. Principles of Expsure Measurement in Epidemiology 2nd Ed Oxford University Press, Oxford UK.

[2] Quinn TC, Wawer MJ, Sewankambo N, et al. Viral load and heterosexual transmission of human immunodeficiency virus type 1. Rakai Project Study Group. *N Engl J Med* 2000; 342: 921–29

[3] Maartens G, Celum C, Lewin SR. HIV infection: epidemiology, pathogenesis, treatment, and prevention. *The Lancet* 2014; 384(9939):258-271.

[4] Mugavero MJ, Napravnik S, Cole SR, Eron JJ, Lau B, Crane HM, et al. Viremia copy-years predicts mortality among treatment-naive HIV-infected patients initiating antiretroviral therapy. *Clinical infectious diseases*. 2011;53(9):927-35.

[5] Olson AD, Walker AS, Suthar AB, Sabin C, Bucher HC, Jarrin I, et al. Limiting Cumulative HIV Viremia Copy-Years by Early Treatment Reduces Risk of AIDS and Death. *Journal of acquired immune deficiency syndromes.* 2016;73(1):100-8.

[6] Chirouze C, Journot V, Le Moing V, Raffi F, Piroth L, Reigadas S, et al. Viremia copy-years as a predictive marker of all-cause mortality in HIV-1-infected patients initiating a protease inhibitor-containing antiretroviral treatment. *Journal of acquired immune deficiency syndromes*. 2015;68(2):204-8.

[7] Zoufaly A, Stellbrink HJ, Heiden MA, Kollan C, Hoffmann C, van Lunzen J, et al. Cumulative HIV viremia during highly active antiretroviral therapy is a strong predictor of AIDS-related lymphoma. *The Journal of infectious diseases*. 2009;200(1):79-87.

[8] Cole SR, Napravnik S, Mugavero MJ, Lau B, Eron JJ, Jr., Saag MS. Copy-years viremia as a measure of cumulative human immunodeficiency virus viral burden. *American journal of epidemiology*. 2010; 171(2): 198 - 205.

[9] Quiros-Roldan E, Raffetti E, Castelli F, Foca E, Castelnuovo F, Di Pietro M, et al. Low-level viraemia, measured as viraemia copy-years, as a prognostic factor for medium-long-term all-cause mortality: a MASTER cohort study. *The Journal of antimicrobial chemotherapy*. 2016;71(12):3519-27.

[10] Wright ST, Hoy J, Mulhall B, O'Connor C C, Petoumenos K, Read T, et al. Determinants of viremia copy-years in people with HIV/AIDS after initiation of antiretroviral therapy. *Journal of acquired immune deficiency syndromes*. 2014;66(1):55-64.

[11] Myer L, Phillips TK, Zerbe A, Ronan A, Hsiao NY, Mellins CA, et al. Optimizing Antiretroviral Therapy (ART) for Maternal and Child Health (MCH): Rationale and Design of the MCH-ART Study. *Journal of acquired immune deficiency syndromes.* 2016;72 Suppl 2:S189-96.

[12] Lefrere JJ, Morand-Joubert L, Mariotti M, Bludau H, Burghoffer B, Petit JC, et al. Even individuals considered as long-term nonprogressors show biological signs of progression after 10 years of human immunodeficiency virus infection. *Blood*. 1997;90(3):1133-40.

[13] Cates JE, Westreich D, Edmonds A, Wright RL, Minkoff H, Colie C, et al. The Effects of Viral Load Burden on Pregnancy Loss among HIV-Infected Women in the United States. *Infectious diseases in obstetrics and gynecology*. 2015;2015:362357.

[14] Lima VD, Sierra-Madero J, Wu Z, Singer J, Wood E, Hull MW, et al. Comparing the efficacy of efavirenz and boosted lopinavir using viremia copy-years. *Journal of the International AIDS Society*. 2014;17:18617.

[15] Kukoyi O, Renner L, Powell J, Barry O, Prin M, Kusah J, et al. Viral load monitoring and antiretroviral treatment outcomes in a pediatric HIV cohort in Ghana. *BMC infectious diseases*. 2016;16:58.

[16] Kowalkowski MA, Day RS, Du XL, Chan W, Chiao EY. Cumulative HIV viremia and non-AIDS-defining malignancies among a sample of HIV-infected male veterans. *Journal of acquired immune deficiency syndromes*. 2014;67(2):204-11.

[17] Atkinson, K.,E. An Introduction to Numerical Analysis (2nd ed.). New York: John Wiley & Sons. 1989.

[18] Sempa JB, Dushoff J, Daniels MJ, Castelnuovo B, Kiragga AN, Nieuwoudt M, et al. Reevaluating Cumulative HIV-1 Viral Load as a Prognostic Predictor: Predicting Opportunistic Infection Incidence and Mortality in a Ugandan Cohort. *American journal of epidemiology.* 2016;184(1):67-77.

[19] Lesosky M, Glass T, Mukonda E, Hsiao NY, Abrams EJ, Myer L. Optimal timing of viral load monitoring during pregnancy to predict viraemia at delivery in HIV-infected women initiating ART in South Africa: a simulation study. *Journal of the International AIDS Society*. 2017;20 Suppl 7.

[20] Gous NM, Berrie L, Dabula P, Stevens W. South Africa's experience with provision of quality HIV diagnostic services. *African Journal of Laboratory Medicine*. 2016;5(2):436. doi:10.4102/ajlm.v5i2.436.

[21] South African National Department of Health. National consolidated guidelines for the prevention of mother to-child transmission of HIV (PMTCT) and the management of HIV in children, adolescents and adults. Pretoria. 2015.

[22] Laut KG, Shepherd LC, Pedersen C, Rockstroh JK, Sambatakou H, Paduta D, et al. Associations between HIV-RNA-based indicators and virological and clinical outcomes. *AIDS* 2016;30(12):1961-72.

[23] Salinas JL, Rentsch C, Marconi VC, Tate J, Budoff M, Butt AA, et al. Baseline, Time-Updated, and Cumulative HIV Care Metrics for Predicting Acute Myocardial Infarction and All-Cause Mortality. *Clinical infectious diseases* 2016;63(11):1423-30.